\documentclass[9pt,conference]{IEEEtran}
\usepackage{dcase2026}

\usepackage[dvipsnames]{xcolor}
\usepackage{bm} % for bold math symbols (incl. Greek letters) with \bm{}
\usepackage[nosfdefault]{comicneue}

\newcommand{\ono}[1]{%
  {\comicneue #1}%
}

\usepackage{dcase2026,amsmath,graphicx,url,times,booktabs, tabularx}
\usepackage{multirow}

\newcommand{\argmax}{\mathop{\rm arg~max}\limits} % Definition \argmax ; \argmax_{x}
\newlength\savedwidth
\newcommand{\wcline}[1]{\noalign{\global\savedwidth\arrayrulewidth\global\arrayrulewidth 1.0pt} \cline{#1}
\noalign{\global\arrayrulewidth\savedwidth}}

\usepackage[most]{tcolorbox}

\newtcolorbox{promptblock}{
    colback=white,
    colframe=black,
    boxrule=0.5pt,
    arc=0pt,
    left=6pt,
    right=6pt,
    top=6pt,
    bottom=6pt,
    width=0.9\linewidth,
    fontupper=\small\ttfamily,
    breakable
}

\title{Misrecognition or Abstraction?\\Rethinking Outputs of Sound Event Recognition}

\name{Naoya Tomida$^{1}$, Yuki Okamoto$^{2}$, and Keisuke Imoto$^{1}$ \vspace{-8pt}
\address{$^{1}$Kyoto University, Japan\ \ \ 
$^{2}$The University of Tokyo, Japan
\thanks{\ \\[-5pt] This work was supported by the Hoso Bunka Foundation, ROIS NII Open Collaborative Research 2026-262S08-24708, and JSPS KAKENHI Grant Number 25K21221.}
}}

\begin{document}

\maketitle

%----------------------------------------------------------------
\begin{abstract}
%----------------------------------------------------------------
Conventional general sound recognition systems typically output deterministic sound event labels, implicitly assuming that the target sound class can be correctly identified from the input audio. However, in real listening situations, the sound event class is not always clearly identifiable. Human listeners may nevertheless understand their surroundings from an ambiguous sound without identifying its exact sound event class. This motivates a discussion of how the outputs of sound recognition systems should be redesigned under such uncertainty. As a basis for this discussion, this paper proposes an output representation for sound event recognition that combines a sound event class, its confidence score, and an onomatopoeic description of the sound. The proposed representation preserves conventional class-based recognition while providing an additional onomatopoeic description of acoustic characteristics that can remain informative even when the class prediction is uncertain. Experiments using ESC-50 and ESC-50-Onomatopoeia show that the proposed method achieves sound recognition performance comparable to that of a conventional recognition-only system. In addition, an LLM-as-a-judge evaluation and subjective listening experiments indicate that the proposed output is preferred over conventional deterministic outputs based on the sound event label, particularly when used to support understanding of the surrounding environment. These results suggest that such output representations can make sound event recognition more informative and communicative under uncertainty.
%----------------------------------------------------------------
\end{abstract}
%----------------------------------------------------------------
%
%----------------------------------------------------------------
%----------------------------------------------------------------
%
%----------------------------------------------------------------
\section{Introduction}
\label{sec:intro}
%----------------------------------------------------------------
General sound recognition, which aims to analyze and understand a wide range of sounds beyond speech and music, has become an important research topic in computational audio understanding.
It has been explored in applications such as machine condition monitoring \cite{Koizumi_DCASE2020_01}, ecological assessment \cite{Acevedo_EI2009_01}, multimedia retrieval \cite{Wold_MultiMedia1996_01}, and home monitoring \cite{Imoto_INTERSPEECH2013_01}.

Common tasks in this field include sound event recognition (SER), audio tagging \cite{Foster_WASPAA2015_01}, and sound event localization and detection (SELD) \cite{Adavanne_JSTSP2019_01}.
In these tasks, systems primarily predict the sound event classes contained in the input audio, with some tasks also estimating when and where the sounds occur.
Beyond these basic tasks, automated audio captioning \cite{Drossos_WASPAA2017_01} and audio question answering \cite{Abdelnour_NeurIPS2018_01} have also been studied to produce more descriptive or interactive outputs.
Rather than outputting sound event labels, these tasks generate natural-language descriptions of sound events or answers to questions about the input audio.
Even in these tasks, however, the outputs are still grounded in the recognition of sound events, together with their temporal order, co-occurrence, and surrounding context.

Existing work on general sound recognition has largely assumed that systems should correctly identify the sound event class in the input audio.
When this identification fails, the output is simply regarded as a recognition error to be reduced by improving the model architecture, acoustic features, or training data.
In human listening, however, assigning an exact category name is not always necessary for using a sound to understand the surrounding situation.
For example, a person may hear a loud explosive sound while its exact source remains unclear. 
Despite this uncertainty, the person still understands the sound through an imitative expression like ``a loud BANG, like a small explosion or something being fired,'' prompting them to check the surroundings or warn others nearby.
This suggests that general sound recognition systems should be designed under the assumption that the sound event class cannot always be correctly identified, and that their output representations should be redesigned accordingly.

Following this view, we propose to represent the output of a sound event recognition system as a combination of a sound event class, its confidence score, and an onomatopoeic description of the sound.
This output is intended to preserve the conventional class-based prediction while also providing a description that remains useful when the prediction is uncertain.
As a simple implementation, we adapt contrastive language-audio pretraining (CLAP) \cite{Wu_ICASSP2023_01} to onomatopoeic expressions and use it to perform sound recognition and onomatopoeia retrieval jointly.
Experimental results show that the proposed method achieves sound event recognition performance comparable to that of a conventional recognition-only system.
In addition, an LLM-as-a-judge evaluation and subjective listening experiments show that the proposed output is preferred over conventional outputs for systems that support surrounding-environment understanding.

The remainder of this paper is organized as follows.
Section 2 reviews conventional sound recognition systems and the general ideas behind their outputs.
Section 3 then describes the proposed output representation for sound event recognition, along with a representative implementation.
Section 4 presents the fundamental SER performance of the proposed method and the results of an LLM-as-a-judge evaluation and subjective evaluations, followed by a discussion on desirable output representations for sound recognition systems.
Finally, Section 5 concludes the paper.
\begin{figure}[t]
\centering
\vspace{0pt}
\includegraphics[width=0.91\columnwidth]{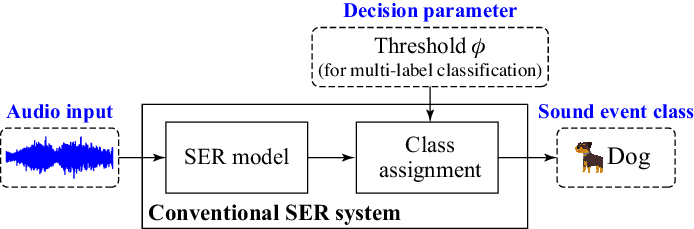}
\vspace{-7pt}
\caption{Conventional output of SER system}
\label{fig:conventional_output}
%\vspace{0pt}
\end{figure}
\begin{figure}[t]
\vspace{-5pt}
\centering
\includegraphics[width=0.93\columnwidth]{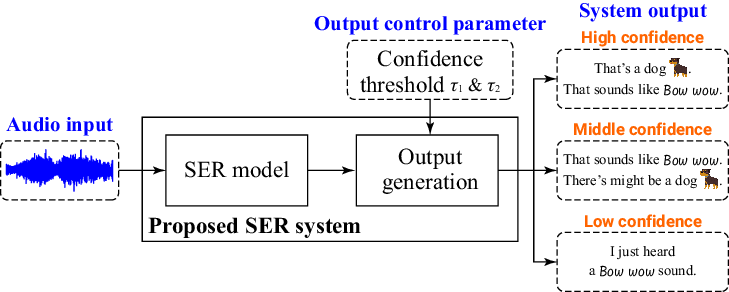}
\vspace{-7pt}
\caption{Example of proposed output representation of SER system}
\label{fig:proposed_output}
\vspace{3pt}
\end{figure}
\begin{figure*}[t]
\vspace{-2pt}
\centering
\includegraphics[width=1.53\columnwidth]{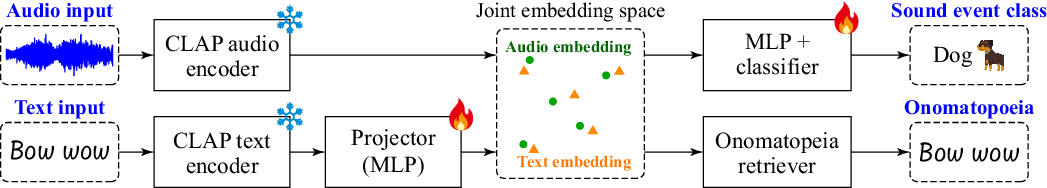}
\vspace{-6pt}
\caption{Overview of implementation of proposed method}
\label{fig:implementation}
\vspace{3pt}
\end{figure*}
%
%
%----------------------------------------------------------------
\section{Conventional Output Representations of Sound Event Recognition}
%----------------------------------------------------------------
In this section, we provide a brief overview of typical implementations of sound recognition systems, with a focus on how outputs are commonly provided in conventional approaches, as shown in Fig.~\ref{fig:conventional_output}.
In many sound recognition pipelines, the input acoustic signal is first transformed into time--frequency representations $\mathbf{X} \in \mathbb{R}^{D \times T}$, where $D$ and $T$ represent the feature dimension along the frequency axis and the number of time frames, respectively, typically using spectrograms or log mel-spectrograms.
The extracted feature $\mathbf{X}$ is then fed into an SER system $f$, which outputs a logit $\mathbf{y} = f(\mathbf{X}) \in \mathbb{R}^{C}$, where $C$ denotes the number of sound event classes.
The logit $\mathbf{y}$ is subsequently converted into class scores by applying an activation function, such as the softmax function for single-label classification or the sigmoid function for multi-label classification.
The final sound event prediction is obtained through class assignment, which selects the sound event class with the highest score in the single-label setting.
In multi-label classification, the assignment instead selects all classes whose scores exceed a threshold $\phi$ as follows:

\vspace{-6pt}
\begin{align}
\!\!\! \hat{\mathcal{C}} =
\begin{cases}
\displaystyle \argmax_{c} \ \mathrm{softmax}_{c} (\mathbf{y}) & \text{(single-label classification)} \\
\{\, c \mid \sigma(y_{c}) \ge \phi \,\} & \text{(multi-label classification)}
\end{cases}
\end{align}
\vspace{-2pt}

\noindent where $c$ is the index of the sound event class, $y_{c}$ is the $c$-th element of $\mathbf{y}$, $\mathrm{softmax}_{c}$ denotes the $c$-th element of the softmax output, $\sigma$ is the sigmoid function, and $\phi$ is the detection threshold.

This formulation indicates that conventional sound recognition systems produce deterministic and discrete sound event labels.
The system output is therefore reduced to a decision about which sound event class is present, while other information is discarded.
One possible way to retain information beyond the decided class label is to use the scores obtained after applying the softmax or sigmoid function as a proxy for prediction confidence.
Although they can reflect relative confidence, they are not necessarily calibrated probabilities of correctness \cite{Guo_ICML2017_01}.

Meanwhile, imitative expressions such as onomatopoeic expressions offer another way to describe the input sound when the sound event class is not clearly identifiable.
Several methods have exploited such a relationship between sounds and onomatopoeic expressions in audio retrieval \cite{Ikawa_DCASE2018_01}, audio-to-onomatopoeia conversion \cite{Miyazaki_EUSIPCO2018_01,Kim_CHI2025_01}, and sound generation \cite{Okamoto_ATSIP2021_01}.
Nevertheless, limited attention has been paid to how SER systems should handle misrecognitions or how outputs should be presented when the confidence is low.
%
%
%----------------------------------------------------------------
\section{Proposed Output of Sound Event Recognition}
%----------------------------------------------------------------
%- - - - - - - - - - - - - - - - - - - - - - - - - - - - - - - - - -
\subsection{Motivation and strategy}
%- - - - - - - - - - - - - - - - - - - - - - - - - - - - - - - - - -
As described in Section 2, conventional sound recognition systems are designed to produce deterministic and discrete sound event labels, regardless of the reliability of the prediction.
This design raises a fundamental question about what the SER system should output when the prediction is unreliable.
In human perception and description of sounds, the form of expression often depends on the confidence level.
When the sound source is clearly identifiable, explicit sound event labels are used.
In contrast, when the sound is ambiguous or the confidence is low, descriptions tend to rely on acoustic characteristics such as temporal and spectral patterns, often expressed using imitative expressions such as onomatopoeia \cite{Lemaitre_JASA2014_01}.
In this work, we thus introduce an alternative output representation that combines a sound event label, its confidence score, and an onomatopoeic description of the input sound, as shown in Fig.~\ref{fig:proposed_output}.
Such an output is expected to better align with how humans adapt their descriptions of sounds according to their confidence.
Moreover, comparing the conventional and proposed output representations enables an investigation of how the output should be designed when the confidence is low.
%
%
%- - - - - - - - - - - - - - - - - - - - - - - - - - - - - - - - - -
\subsection{Implementation of proposed method}
%- - - - - - - - - - - - - - - - - - - - - - - - - - - - - - - - - -
There are several ways to implement the proposed framework, such as jointly modeling sound event recognition with reliability estimation and sequence-to-sequence transcription from audio to text.
In this work, we adopt a simple approach that jointly performs sound event recognition, confidence estimation, and onomatopoeia retrieval from audio input.

An overview of our implementation is illustrated in Fig.~\ref{fig:implementation}.
In this study, we use a shared embedding space between audio and onomatopoeic text representations, based on pretrained CLAP embeddings \cite{Wu_ICASSP2023_01}.
CLAP provides a joint audio-language embedding space; however, the existing CLAP models are not trained with sufficient amounts of onomatopoeic text.
As a result, audio and onomatopoeic text embeddings may not be sufficiently aligned in the pretrained embedding space.
To address this issue, we introduce a simple projector implemented as an MLP applied to the text embeddings produced by the CLAP text encoder.
The projector is trained to better align the embeddings of onomatopoeic text with the corresponding audio embeddings.

During training, the audio and text encoders are frozen, and the projector is trained to improve the alignment using the following objective:

\vspace{-17pt}
\begin{align}
  \mathcal{L}_{\mathsf{ono}} = \frac{1}{N} \sum_{i=1}^{N} \left\| \mathbf{a}_{i} - g(\mathbf{t}_{i}) \right\|^{2}_{2},
\end{align}
\vspace{-7pt}

\noindent where $\mathbf{a}_{i}$ and $\mathbf{t}_{i}$ indicate the $i$-th audio and onomatopoeic text embeddings, respectively.
$g$ is the projector and $N$ is the number of training pairs.

In addition, to predict sound event classes from the audio embeddings, we jointly train a classifier composed of several MLP layers.
Although this classifier can be replaced with predictors tailored to various downstream tasks, such as SED, SELD, or audio captioning, we focus on a simple sound event classification as an initial step in this study.
The sound event classifier is trained using the following objective:

\vspace{-17pt}
\begin{align}
\mathcal{L}_{\mathsf{cls}} = - \sum_{c=1}^{C} z_{c} \log \mathrm{softmax}_{c} (\mathbf{y}),
\end{align}
\vspace{-7pt}

\noindent where $z_{c}$ is the ground-truth label for sound event class $c$.
The overall model is trained with the following objective:

\vspace{-10pt}
\begin{align}
\mathcal{L} = \mathcal{L}_{\mathsf{ono}} + \lambda \mathcal{L}_{\mathsf{cls}},
\end{align}
\vspace{-7pt}

\noindent where $\lambda$ is a weighting hyperparameter.

At inference time, the proposed system jointly produces a sound event prediction and an onomatopoeic description from the input audio.
The trained sound event classifier outputs the predicted sound event class with its confidence score, defined in our implementation as the margin-based confidence score \cite{Scheffer_IDA2001_01}, computed as the normalized difference between the top-two softmax scores:

\vspace{-10pt}
\begin{align}
s = \frac{p_{(1)} - p_{(2)}}{p_{(1)} + p_{(2)}},
\end{align}
\vspace{-7pt}

\noindent where $p_{(1)}$ and $p_{(2)}$ are the softmax probabilities of the top-two sound event classes.
Note that other confidence measures can be used instead.
Meanwhile, the onomatopoeic description is retrieved from the candidate set using the CLAP embedding space.
Specifically, the system retrieves the onomatopoeia whose projected text embedding has the smallest Euclidean distance to the audio embedding:

\vspace{-10pt}
\begin{align}
\hat{o} = \arg\min_{o \in \mathcal{O}} \left\| \mathbf{a} - g(\mathbf{t}_{o}) \right\|_{2},
\end{align}
\vspace{-7pt}

\noindent where $\mathcal{O}$ denotes the set of candidate onomatopoeic expressions, $\mathbf{a}$ is the audio embedding, and $\mathbf{t}_{o}$ is the text embedding of onomatopoeic expression $o$.
The final output form is then determined according to the confidence score of the predicted sound event class, as shown in Fig.~\ref{fig:proposed_output}.
%
%
%- - - - - - - - - - - - - - - - - - - - - - - - - - - - - - - - - -
\vspace{-2pt}
\subsection{Output design for sound recognition results}
\vspace{-2pt}
%- - - - - - - - - - - - - - - - - - - - - - - - - - - - - - - - - -
The proposed output representation enables the system to adjust its output form according to the confidence of the sound event prediction.
This is expected to prevent the system from asserting incorrect sound event labels when the prediction is unreliable, and to provide useful information to the user even in such cases.

One simple implementation is to switch between predefined output templates according to the prediction confidence, as shown in Fig.~\ref{fig:proposed_output}.
For example, when the input contains a dog barking sound, the system may output ``That is a dog barking sound, and the sound is like bow wow.'' if the confidence score of the predicted class is greater than or equal to 0.5.
If the confidence score is between 0.2 and 0.5, the system may instead output ``That sound is like bow wow, and that might be a dog barking sound.''
If the confidence score is below 0.2, the system may output only ``I heard a bow wow like sound.''
This template-based design is easy to implement, but its outputs can become repetitive and may be perceived as formulaic and impersonal.

A more flexible design is possible when the SER system is connected to an LLM-based dialogue system or agent.
In this case, the predicted sound event label, its confidence score, and the onomatopoeia can be provided to the LLM to generate a natural response.
\begin{table}[t]
\footnotesize
\caption{Detailed experimental conditions}
\label{tbl:parameter}
\centering
\begin{tabular}{ll}
\wcline{1-2}
&\\[-7pt]
Network for MLP and classifier&3 dense layers\\[1pt]
\# units in MLP and classifier layers&512, 256, 128\\[0pt]
$\lambda$&1.0\\[0pt]
Optimizer&AdamW \cite{Loshchilov_ICLR2019_01}\\[0pt]
\wcline{1-2}
\end{tabular}
\vspace{6pt}
\end{table}
%
%
%----------------------------------------------------------------
\section{Evaluation Experiments}
%----------------------------------------------------------------
%- - - - - - - - - - - - - - - - - - - - - - - - - - - - - - - - - -
\subsection{Experimental conditions}
%- - - - - - - - - - - - - - - - - - - - - - - - - - - - - - - - - -
We conducted evaluation experiments to assess the fundamental SER performance of the proposed method, as well as to examine whether the proposed output representation is suitable for conveying the recognition results through a conversational robot.
For evaluation, we used the ESC-50 dataset \cite{Piczak_ACMMM2015_01} as the audio data and ESC-50-Onomatopoeia\footnote{\url{https://github.com/Y-Okamoto1221/ESC-50-Onomatopoeia}} as the onomatopoeic text data.
The ESC-50-Onomatopoeia dataset provides onomatopoeic annotations for a subset of 31 sound classes from ESC-50.
In total, the dataset consists of 1,240 audio samples (31 sound classes $\times$ 40 samples) and 186,000 onomatopoeic descriptions (31 sound classes $\times$ 40 audio samples $\times$ 150 onomatopoeia), all used as individual audio-onomatopoeia pairs.

The audio input was represented using 64-dimensional log-mel spectrograms extracted from audio sampled at 16 kHz, with a window size of 1024 and a hop length of 480.
For the audio encoder, we employed the Hierarchical Token Semantic Audio Transformer (HTS-AT) \cite{Chen_ICASSP2022_01}, while RoBERTa \cite{Liu_arXiv2019_01} was used as the text encoder.
Both encoders were used with the pretrained CLAP checkpoint \texttt{630k-audioset-best.pt (non-fusion)} \cite{Wu_ICASSP2023_01}.
The other experimental conditions are shown in Table~\ref{tbl:parameter}.
%
%
%- - - - - - - - - - - - - - - - - - - - - - - - - - - - - - - - - -
\subsection{Sound event recognition performance}
%- - - - - - - - - - - - - - - - - - - - - - - - - - - - - - - - - -
We first examined whether the proposed method preserves the fundamental sound event recognition capability.
For comparison, we also evaluated a conventional SER system that uses the same audio encoder and classification head as those used in the proposed method.
This baseline excludes the CLAP text encoder, the projector, and the onomatopoeia retrieval module shown in Fig.~\ref{fig:implementation}, and is trained only with $\mathcal{L}_{\mathsf{cls}}$.
The experiments followed the official five-fold cross-validation setup of ESC-50 \cite{Piczak_ACMMM2015_01}, applied to the 31-class subset covered by ESC-50-Onomatopoeia, and each fold configuration was repeated ten times with different random seeds.
All hyperparameters were empirically determined in advance and were not tuned using the evaluation split.

Table~\ref{tbl:ser_performance_01} summarizes the sound event recognition results.
The reported values are the means and standard deviations over the five folds and ten random seeds.
The proposed system achieved recognition performance comparable to that of the conventional baseline.
These results indicate that the additional training for audio--onomatopoeia alignment does not compromise the sound event recognition capability of the system.
\begin{table}[t]
\footnotesize
\caption{SER performance on 31-class subset of ESC-50 (\%)}
\label{tbl:ser_performance_01}
\centering
\begin{tabular}{lrr}
\wcline{1-3}
&\\[-6pt]
\multicolumn{1}{c}{\bf{Method}}&\multicolumn{1}{c}{\bf{Micro-F score}}&\multicolumn{1}{c}{\bf{Macro-F score}}\\
\cline{1-3}
&\\[-6pt]
CLAP + classifier&97.81 $\pm$ 0.66&97.80 $\pm$ 0.66\\[0pt]
CLAP + onoma.&\multirow{2}{*}{98.54 $\pm$ 0.51}&\multirow{2}{*}{98.52 $\pm$ 0.51}\\[0pt]
(Proposed)&\\[0pt]
\wcline{1-3}
\end{tabular}
\vspace{6pt}
\end{table}
\begin{figure}[t]
\centering
\setlength{\fboxsep}{3pt}
\setlength{\fboxrule}{0.1pt}
\fbox{
\begin{minipage}{0.46\textwidth}
\fontsize{6.2pt}{6.2pt}\selectfont
\ttfamily
You are an evaluator of the output representations of a sound event recognition (SER) system. The system is connected to a human-like conversational robot that perceives surrounding sounds and speaks to a nearby user based on the system output.
Suppose that you have received two recognition results A and B for the same sound. Rate each result and judge which one is preferable and give a mean opinion score (MOS) with a five-point scale (1: very poor - 5: excellent) to each result. Also provide the reason for your judgment.\\[-1.8pt]

\textit{[The remainder of the prompt is omitted for brevity.]}
\end{minipage}
}
\caption{Prompt used for the LLM-as-a-judge evaluation. The SER system outputs, ground-truth sound event labels, and instructions specifying the required evaluation output format are omitted for brevity.}
\label{Fig:llm_prompt}
\vspace{6pt}
\end{figure}
\begin{table*}[t]
\vspace{-3pt}
\footnotesize
\caption{Examples of system outputs under high- and low-confidence predictions. Onomatopoeias are written in an IPA-like format.}
\label{tbl:output_example}
\centering
\begin{tabular}{cccc}
\wcline{1-4}
&\\[-6pt]
\multicolumn{1}{c}{\bf{Ground-truth}}&\multicolumn{1}{c}{\bf{CLAP + classifier}}&\multicolumn{1}{c}{\bf{CLAP + classifier}}&\multicolumn{1}{c}{\bf{CLAP + onoma.}}\\[-1pt]
\multicolumn{1}{c}{\bf{label}}&\multicolumn{1}{c}{\bf{(event label)}}&\multicolumn{1}{c}{\bf{(LLM)}}&\multicolumn{1}{c}{\bf{(template)}}\\
\cline{1-4}
&\\[-7pt]
\bf{cow} & \textcolor{blue}{\bf{cow}} & Did you hear that, too? It sounded like \textcolor{blue}{\bf{cow}}. & I heard a "\ono{\textcolor{RedOrange}{m o o:}}" sound, which sounded like \textcolor{blue}{\bf{cow}}.\\[2pt]
\multirow{2}{*}{\bf{engine}} & \multirow{2}{*}{\textcolor{blue}{\bf{chainsaw}}} & Something just made a sound around us,& I heard a sound nearby, but I can only \\[-1pt]
& & and I think it may have been \textcolor{blue}{\bf{chainsaw}}. & describe it as something like "\ono{\textcolor{RedOrange}{g a g a g a g a g a}}".\\[0pt]
&\\[-8pt]
\wcline{1-4}
\end{tabular}
\vspace{1pt}
\end{table*}
%
%
%
%- - - - - - - - - - - - - - - - - - - - - - - - - - - - - - - - - -
\subsection{LLM-as-a-judge evaluation of output representations}
%- - - - - - - - - - - - - - - - - - - - - - - - - - - - - - - - - -
We then evaluated whether the proposed output representation for sound event recognition is preferable to the conventional outputs when the SER system is used as a front-end for spoken dialogue systems.
As a first evaluation, we used an LLM-as-a-judge framework (with ChatGPT 5.5 middle) to assess the system outputs through (i) pairwise comparisons between the two output formats and (ii) mean opinion score (MOS) ratings in terms of utterance naturalness, richness of sound description, and communicative usefulness.

In this experiment, we assumed a scenario in which the SER system is connected to a conversational robot that operates in the real world, autonomously perceives its surrounding acoustic environment, and proactively initiates communication with users.
Under this assumption, the LLM judge rated the system outputs on the three criteria above.
As illustrated in Fig.~\ref{Fig:llm_prompt}, the LLM judge was provided with a description of the assumed situation, the sound recognition results with their confidence scores and onomatopoeias, and the required format of the evaluation output.
Note that the input audio itself was not provided to the judge.

As baselines for comparison, we prepared two output representations based on the conventional CLAP-based deterministic SER system.
The first baseline directly uses the predicted sound event label as the system output.
In the second baseline, the predicted sound event label is fed into an LLM (ChatGPT 5.5 middle) and reformatted into a natural utterance.
Note that, as shown in Table~\ref{tbl:ser_performance_01}, both the conventional and proposed systems achieve very high recognition accuracy on the ESC-50 subset, which makes it difficult to assess the quality of the output representations in the presence of misrecognitions or low-confidence predictions.
To emulate uncertain predictions, in this experiment, we selected the number of training epochs so that the recognition accuracy was approximately 85\% for both systems, which also degraded the audio--onomatopoeia alignment of the proposed system.
Examples of the system outputs generated by each method for high- and low-confidence predictions are shown in Table~\ref{tbl:output_example}.

\begin{figure}[t]
\centering
\includegraphics[width=1.0\columnwidth]{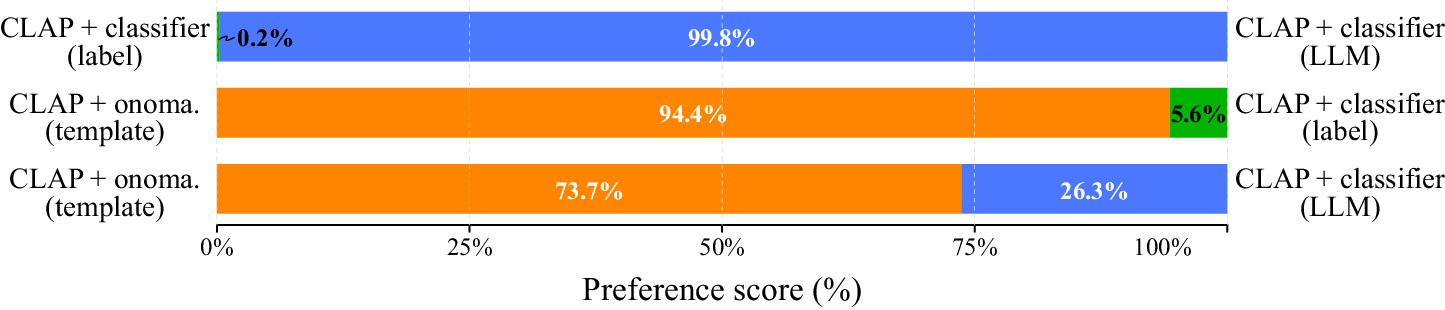}
\vspace{-17pt}
\caption{Pairwise preference scores for SER output representations, obtained using LLM-as-a-judge}
\label{fig:preference_score_01}
\vspace{-2pt}
\end{figure}

Figure~\ref{fig:preference_score_01} and Table~\ref{tbl:MOS_01} show the pairwise comparison and MOS evaluation results, respectively. 
Overall, both results indicate that the outputs of the proposed system were preferred over those of the conventional deterministic systems across all evaluation criteria, with a particularly substantial advantage in richness of sound description.

In addition, the LLM judge was instructed to provide the reasons for its judgments, and we examined them to identify the benefits and limitations of the proposed output representation.
In terms of utterance naturalness, the proposed outputs were positively evaluated for naturally expressing uncertainty in the recognition results. However, excessively long or repetitive onomatopoeic expressions sometimes made the resulting utterances awkward and less natural.
Regarding richness of sound description, the proposed system was considered capable of providing more concrete and detailed descriptions of the perceived sounds.
In contrast, when the onomatopoeia did not appropriately describe the input sound, it degraded rather than enriched the sound description.
For communicative usefulness, explicitly conveying the recognition uncertainty was regarded as useful information for users.
Nevertheless, outputs that did not provide any candidate sound event label under low-confidence conditions were sometimes considered less useful.

These findings indicate that further improvements in both SER accuracy and onomatopoeia retrieval (or generation) accuracy are necessary.
The findings also suggest that a more sophisticated design of the output representation could further improve the naturalness, descriptive richness, and communicative usefulness of the proposed output representation.
%
%
%- - - - - - - - - - - - - - - - - - - - - - - - - - - - - - - - - -
\subsection{Subjective listening evaluation of system outputs}
%- - - - - - - - - - - - - - - - - - - - - - - - - - - - - - - - - -
We next conducted an MOS evaluation using a crowdsourcing platform with 310 participants, in which the evaluators listened to the input audio and then rated the corresponding system output.
In this experiment, the same three criteria as in the LLM-as-a-judge evaluation were applied.
We evaluated the system outputs for 1,240 samples for each of the three methods.
We excluded 3.2\% of the responses through a screening process, as they contained clearly erroneous answers.
Note that the proposed method was evaluated with the simplest template-based implementation under the intentionally degraded alignment condition.

As shown in Table~\ref{tbl:MOS_02}, the results for richness and usefulness exhibited trends similar to those observed in the LLM-as-a-judge evaluation.
However, the differences in the mean scores among all three methods were smaller, while the variability of the human ratings was greater.
In contrast, little difference was observed among the three methods in terms of utterance naturalness.

\begin{table}[t]
\scriptsize
\caption{Mean LLM-as-a-judge scores for utterance naturalness, richness of sound description, and communicative usefulness. Values are means $\pm$ standard deviations.}
\label{tbl:MOS_01}
\centering
\begin{tabular}{lrrr}
\wcline{1-4}
&\\[-6pt]
\multicolumn{1}{c}{\!\!\bf{Method}}\!\!&\multicolumn{1}{c}{\!\!\bf{Naturalness}\!\!}&\multicolumn{1}{c}{\!\!\bf{Richness}\!\!}&\multicolumn{1}{c}{\!\!\bf{Usefulness}\!\!}\\
\cline{1-4}
&\\[-6pt]
CLAP + classifier (event label)&2.91 $\pm$ 0.31&1.78 $\pm$ 0.47&3.96 $\pm$ 1.08\\[0pt]
CLAP + classifier (LLM)&3.71 $\pm$ 0.55&2.04 $\pm$ 0.25&4.06 $\pm$ 0.27\\[0pt]
CLAP + onoma. (template)&\bf{3.87 $\pm$ 0.36}&\bf{3.97 $\pm$ 0.61}&\bf{4.39 $\pm$ 0.87}\\[0pt]
\wcline{1-4}
\end{tabular}
%\end{table}
%
%
%\begin{table}[t]
\vspace{14pt}
\scriptsize
\caption{Mean opinion scores by human evaluators for utterance naturalness, richness of sound description, and communicative usefulness. Values are means $\pm$ standard deviations.}
\label{tbl:MOS_02}
\centering
\begin{tabular}{lrrr}
\wcline{1-4}
&\\[-6pt]
\multicolumn{1}{c}{\!\!\bf{Method}}\!\!&\multicolumn{1}{c}{\!\!\bf{Naturalness}\!\!}&\multicolumn{1}{c}{\!\!\bf{Richness}\!\!}&\multicolumn{1}{c}{\!\!\bf{Usefulness}\!\!}\\
\cline{1-4}
&\\[-6pt]
CLAP + classifier (event label)&\bf{3.69 $\pm$ 1.05}&2.92 $\pm$ 1.18&3.56 $\pm$ 1.16\\[0pt]
CLAP + classifier (LLM)&3.62 $\pm$ 1.07&3.26 $\pm$ 1.05&3.62 $\pm$ 1.04\\[0pt]
CLAP + onoma. (template)&3.63 $\pm$ 1.02&\bf{3.70 $\pm$ 1.01}&\bf{3.69 $\pm$ 1.05}\\[0pt]
\wcline{1-4}
\end{tabular}
\vspace{8pt}
\end{table}

These results suggest that onomatopoeic expressions that do not appropriately match the perceived impression of a sound can negatively affect the naturalness of the system output. Because the human evaluators, unlike the LLM judge, directly listened to the input audio, such mismatches between the onomatopoeia and the sound were likely more noticeable in this evaluation.
Therefore, the accuracy of onomatopoeia retrieval is crucial, and onomatopoeic expressions should be incorporated into system outputs with care.
Furthermore, the human ratings exhibited greater variability than the LLM-as-a-judge ratings, suggesting substantial individual differences in the preferred form of system outputs.
Such individual differences and audio-aware automatic judges should be investigated in detail in future work.
%
%
%----------------------------------------------------------------
\section{Conclusion}
%----------------------------------------------------------------
This paper reconsidered the output representation of sound event recognition systems, which have conventionally relied on deterministic sound event labels.
To provide more informative outputs when the class prediction is uncertain, we proposed an output representation that combines a sound event class, its confidence score, and an onomatopoeic description of the input sound.
Experiments using ESC-50 and ESC-50-Onomatopoeia showed that the proposed method maintained SER performance comparable to that of a conventional recognition-only system.
LLM-as-a-judge evaluation and subjective listening experiments further indicated that the proposed output was preferred over conventional label-based outputs for supporting surrounding-environment understanding.
These results suggest that the output of sound event recognition should be abstracted according to the confidence of the prediction, rather than always presented as a deterministic sound event label.
\bibliographystyle{IEEEtran}
\bibliography{DCASE2026}
\end{document}